\documentclass[twocolumn,showpacs,preprintnumbers,amsmath,
               amssymb,nofootinbib]{revtex4}

\usepackage{graphicx}
\usepackage{dcolumn}
\usepackage{bm}

\newcommand*{\bea}{\begin{eqnarray}}
\newcommand*{\eea}{\end{eqnarray}}
\newcommand*{\be}{\begin{equation}}
\newcommand*{\ee}{\end{equation}}
\newcommand*{\pd}{\partial}
\newcommand*{\pdm}{\pd_{\mu}}

\newcommand*{\pref}[1]{(\ref{#1})}

\newcommand*{\mn}{{\mu\nu}}

\newcommand*{\nn}{\nonumber}

\newcommand*{\tr}{\mathrm{tr}}

\newcommand*{\indexsep}{,}
\newcommand*{\tl}{\mathrm{tl}}

\begin{document}

\preprint{}

\title{Exploratory study of three-point Green's functions in Landau-gauge Yang-Mills theory}

\author{Attilio Cucchieri}\email{attilio@ifsc.usp.br}
\author{Axel Maas}\email{axelmaas@ifsc.usp.br}
\author{Tereza Mendes}\email{mendes@ifsc.usp.br}
\affiliation{Instituto de F\'\i sica de S\~ao Carlos, Universidade de S\~ao Paulo, \\
              Caixa Postal 369, 13560-970 S\~ao Carlos, SP, Brazil}

\date{\today}

\begin{abstract}
Green's functions are a
central element in the attempt to understand non-perturbative phenomena
in Yang-Mills theory. Besides the propagators, 3-point Green's functions
play a significant role, since they permit access to the running coupling
constant and are an important input in functional methods. Here we present numerical
results for the two non-vanishing 3-point Green's functions in 3d pure
$SU(2)$ Yang-Mills theory in (minimal) Landau gauge, i.e.\ the three-gluon
vertex and the ghost-gluon vertex, considering various kinematical regimes.
In this exploratory investigation the lattice volumes are limited to $20^3$ and $30^3$
at $\beta=4.2$ and $\beta=6.0$.
We also present results for the gluon and the ghost propagators, as well
as for the eigenvalue spectrum of the Faddeev-Popov operator. Finally, we compare
two different numerical methods for the evaluation of the inverse
of the Faddeev-Popov matrix, the point-source and the plane-wave-source methods.
\end{abstract}

\maketitle

\section{Introduction}

The non-perturbative properties of Yang-Mills theory are still an open
and challenging problem, especially the issue of confinement. Nonetheless, much
progress has been made in their understanding over the decades. One
central element in these investigations are Green's functions,
which can describe a quantum field theory completely. In particular, their
infrared behavior has been related to confinement:
in two of the most popular scenarios of confinement, the Gribov-Zwanziger
scenario \cite{Gribov:1977wm,Zwanziger:1993dh,Zwanziger,Zwanziger:2002ia}
and the Kugo-Ojima scenario \cite{Kugo}, a specific behavior is
predicted for the 2-point Green's functions --- the propagators --- in Landau gauge.

According to these scenarios, the Faddeev-Popov ghost propagator should
be enhanced in the infrared limit when compared to a massless-particle pole,
while the gluon propagator should vanish. Calculations using functional
methods agree with these predictions and have found that the propagators
show a power-like behavior in the far infrared region, with characteristic exponents.
These results have been obtained using Dyson-Schwinger equations in
various dimensions \cite{Zwanziger,Alkofer:2000wg,vonSmekal,
Maas:2004se} and have been confirmed by renormalization-group methods \cite{Gies}.
On the other hand, these methods rely on approximations concerning the higher $n$-point
Green's functions, especially the three-point vertices. Thus, it is important
to check explicitly whether the assumptions made for the vertices in these methods
are justified. This has been done, again using functional methods, but as of yet
only for specific momentum configurations of the $n$-point Green's functions and in the
far infrared limit \cite{Alkofer:2004it,Schleifenbaum:2006bq}, or using further approximations
\cite{Schleifenbaum:2004id,Schleifenbaum:2006bq}. In both cases the results are consistent with
the assumptions made so far.

Numerical studies of lattice gauge theories also support the two confinement scenarios
described above, confirming the infrared enhancement of the ghost propagator
\cite{Furui:2003jr,Langfeld,Bloch:2003sk,Sternbeck:2005tk}
and the suppression of the gluon propagator (in the 3d case) at low momenta
\cite{Cucchieri3d,Cucchieri:2003di}.
Let us recall that, in order to probe the infrared limit, one needs to use
very large lattice volumes. Recently, in four dimensions, investigations have been
performed also considering (strongly) asymmetric lattices \cite{Silva}, in
order to have access to small momenta while keeping the lattice volume relatively
small. The results obtained show an infrared suppression of the gluon propagator.
However, these simulations are affected by systematic effects
\cite{Ilgenfritz}, making it difficult to extract quantitative information from this
type of lattices.

Let us stress that the agreement between the functional methods and the lattice
data is still at the qualitative level. In particular, it is still not clear
if the numerical data support a gluon propagator vanishing in the infrared
\cite{Cucchieri:2003di,Boucaud:2006pc}.
Recent studies by functional methods \cite{Fischer:2005ui} suggest nontrivial
effects related to the use of
discretized space-time on compact manifolds,
which could be responsible for this discrepancy.
Let us also recall that, in the continuum, a finite (non-zero) gluon propagator
at zero momentum seems to be compatible with the Gribov-Zwanziger scenario
only in the case of an infrared enhanced ghost-gluon vertex \cite{Boucaud:2005ce,Lerche:2002ep}.
On the other hand, such an enhancement would be at variance with results from functional
methods \cite{Alkofer:2004it,Schleifenbaum:2006bq,Schleifenbaum:2004id} and from recent lattice studies
\cite{Cucchieri:2004sq,Ilgenfritz:2006gp}.

In this work we present an exploratory study of the Landau gauge three-gluon vertex
and of the ghost-gluon vertex for various momentum configurations, in the
three-dimensional pure $SU(2)$ case. After this exploratory study we will be able
to consider (for the interesting cases) very large
lattice volumes and to probe the limit of small momenta.
This is the first numerical study of these vertices in the three-dimensional
case.
Clearly, the 3d simulations are computationally less demanding than those in
4d and they may help to get a better understanding of the 
four-dimensional case.
Furthermore, results in three dimensions are of interest in their own right,
as three-dimensional Yang-Mills theory is the (most relevant part of the) infinite-temperature
limit of its four-dimensional counterpart \cite{Maas}. In addition, a connection exists between 4d Yang-Mills theory in Coulomb gauge and 3d Yang-Mills theory in Landau gauge \cite{Schleifenbaum:2006bq,Zwanziger:2003de}.

Let us note that previous lattice studies of these vertices in the 4d case
\cite{Parrinello:1994wd,Alles:1996ka,Cucchieri:2004sq,Boucaud:2000ey}
usually focused on specific momentum configurations,
mostly with the aim of extracting the running coupling constant and for comparison to
perturbative studies. In this preliminary work we use several kinematical
configurations in order to test the numerical methods employed and to study
the influence due to discretization and to finite-volume effects.
In addition, we present results for the propagators and for the spectrum of
the Faddeev-Popov operator, which plays an important role in the Gribov-Zwanziger
scenario \cite{Gribov:1977wm,Zwanziger:1993dh,Zwanziger,Zwanziger:2002ia,
Sternbeck:2005vs}.

In Section \ref{simpl} the technical details for the generation of the configurations,
the gauge-fixing procedure and the error analysis are given. We report definitions
and results for the propagators in Section \ref{sprop} and for the vertices
in Section \ref{svert}. A summary and outlook conclude this work in Section \ref{ssum}.

\section{Generation of configurations}\label{simpl}

The action considered is the usual, unimproved Wilson action \cite{Montvay:1994cy}
for the $SU(2)$ gauge group.
Configurations are generated using a hybrid-over-relaxation (HOR) update,
consisting of five over-relaxation \cite{Adler} and one heat-bath sweeps.
For the heat-bath update, a mixed Creutz \cite{Creutz:1980zw} and Kennedy-Pendleton
\cite{Kennedy:1985nu} algorithm is employed. The lattices have volumes $V=N^3=20^3$
and $30^3$, and calculations have been performed at $\beta=4.2$ and $\beta=6.0$.
We use 200 HOR updates for thermalization and 40 or 45 HOR
updates between evaluation of the Green's functions.
The results have been obtained in multiple independent runs using hot initial configurations.

For the extraction of the vertices, large statistics have been necessary.
Table \ref{conf} lists the precise values. Also, the expectation value of the
plaquette is given for each lattice volume and $\beta$ value.
The results are in agreement with \cite{Cucchieri:2003di} and \cite{Teper}.
The error on the plaquette is the statistical error including a correlation-time analysis.
We always find that the integrated auto-correlation time \cite{sokal}
is less than 1 HOR sweep. Following Ref.\ \cite{Cucchieri:2003di} we also evaluate
the inverse lattice spacing (in GeV).

\begin{table}
\caption{\label{conf} For each lattice volume $V$ and coupling $\beta$ we report
here the average value of the plaquette $<P>$, the inverse lattice spacing $1/a$ (in GeV),
the number of independent runs and the number of configurations considered
for the evaluation of propagators and vertices. The values of the plaquette, the
data for the propagators and the eigenvalue spectrum (reported in Section \ref{sprop}) have been
obtained using the data set labeled {\em Propagators}.
On the other hand, the propagators involved in the determination of the vertices have
been determined considering the same data set as the corresponding vertex.
Finally, note that the set labeled {\em Three-gluon vertex} includes
all the configurations considered in the other two data sets.
}
\begin{ruledtabular}
\vspace{1mm}
\begin{tabular}{|c|c|c|c|c|c|}
       &         &  & & \multicolumn{2}{c|}{\em Propagators} \cr
\hline
Volume & $\beta$ & $<P>$ & $1/a$ [GeV] & Runs & Config.\ \cr
\hline
$20^3$ & 4.2 & 0.741865(5) & 1.136(8) & 19 & 6161 \cr
\hline
$30^3$ & 4.2 & 0.741860(2) & 1.136(8) & 53 & 10229 \cr
\hline
$20^3$ & 6.0 & 0.824781(3) & 1.733(8) & 18 & 5777 \cr
\hline
$30^3$ & 6.0 & 0.824781(1) & 1.733(8) & 48 & 10099 \\[1mm]
\hline
       &         & \multicolumn{2}{c|}{\em Ghost-gluon vertex} & \multicolumn{2}{c|}{\em Three-gluon vertex} \cr
\hline
Volume & $\beta$ & Runs & Config.\ & Runs & Config.\ \cr
\hline
$20^3$ & 4.2 & 19 & 6903 & 38 & 13064 \cr
\hline
$30^3$ & 4.2 & 36 & 11052 & 89 & 21281 \cr
\hline
$20^3$ & 6.0 & 17 & 7004 & 35 & 12781 \cr
\hline
$30^3$ & 6.0 & 16 & 11172 & 64 & 21271 \cr
\end{tabular}
\end{ruledtabular}
\end{table}

The gauge fixing to Landau gauge has been performed using a stochastic-over-relaxation
method \cite{Cucchieri} with a self-adapting
acceptance probability. The condition for gauge fixing has been a test
on the quantity \cite{Cucchieri}
\bea
e_6&=&\frac{1}{d}\sum_\mu\frac{1}{N_\mu}\sum_c\frac{1}{[\tr(Q_\mu\sigma_c)]^2}\nn\\
&&\quad\qquad\times\sum_{x_\mu}(\tr\{[q_\mu(x_\mu)-Q_\mu]\sigma_c\})^2,
\eea
which was required to be less than $10^{-12}$. Here we use the definitions
\bea
q_\mu(x_\mu)&=&\frac{1}{2i}\sum_{x_\nu,\nu\neq\mu}\big[g(x)U_\mu(x)g(x+e_\mu)^\dagger\nn\\
&&\qquad\qquad -g(x+e_\mu)U_\mu(x)^\dagger g(x)^\dagger \big] \label{eq:defqmu} \\
Q_\mu&=&\frac{1}{N_\mu}\sum_{x_\mu}q_\mu(x_\mu)\label{gfq} .
\eea
Also, 
$\{ U_{\mu}(x) \}$ is a thermalized lattice configuration,
$\{ g (x) \}$ represents the gauge transformation applied on
the link variables $ U_{\mu}(x) $, the symbol $^\dagger$ indicates Hermitian
conjugate, $ N_{\mu} $ is
the lattice side in the $\mu$ direction, $d$ is the space-time
dimension ($3$ in our case), $e_{\mu}$ is a
positive unit vector in the $\mu$ direction and $\sigma_{c}$
are the three Pauli matrices, normalized as $\sigma_c^2=1$.
Both the link variables $ U_{\mu}(x) $ and the
gauge-transformation matrices $ g (x) $ are elements of
the $SU(N_c)$ group (in the fundamental $N_c \times N_c$ representation).
In order to speed up the calculation, the quantity $e_6$ was not evaluated for
each stochastic-over-relaxation sweep, but an adaptive-predictor
method has been used for each configuration. Thus, in some cases
we obtain for $e_6$ a value considerably smaller 
(by some orders of magnitude) than $10^{-12}$.
Nevertheless, we usually found for the quantity $e_6$ a final value
between $10^{-12}$ and $10^{-13}$.
No results on Gribov-copy effects will be given in this exploratory study,
but tests using relatively small statistics, i.e.\ generating 20 Gribov
copies for each configuration and considering only a small subset of thermalized
configurations,
suggest that the effects are at the quantitative rather than at the qualitative level.

All errors given for the propagators and the vertices have been calculated
using a standard bootstrap method with 1000 bootstrap samples.
The quoted errors represent a $67\%$ confidence interval.

Most of the results shown here have been obtained using a code developed independently
from the one used in Ref.\ \cite{Cucchieri:2003di} (see also \cite{Cucchieri:2003zx}).
Nonetheless, when possible, we checked that results obtained with the two codes agree for propagators
and vertices.

\section{Propagators}\label{sprop}

\subsection{Gluon propagator}
\label{gluonprop}

The gluon propagator is given by the correlation function (see for example \cite{Bloch:2003sk})
\be
D_\mn^{ab}(p)=\frac{1}{V}<A_\mu^a(p)A_\nu^b(-p)>,\label{gpbas}
\ee
with the momentum-space lattice gluon field defined as \cite{Alles:1996ka}
\be
A_\mu^a(p)=e^{- \frac{i\pi p_{\mu}}{N}}
   \sum_x \frac{e^{2\pi i px/N}}{4i}\tr\left[\left(U_\mu(x)-U_\mu(x)^{+}\right)\sigma_a\right] .
\label{eq:Aofp}
\ee
Here the components $p_\mu$ of $p$ get the integer values
$-N/2 + 1\, , \ldots , \, N/2 \, $ (for even lattice sides
and for symmetric lattices with $N_\mu = N$).
Note that we do not divide the Fourier transform of the gluon field by the
lattice volume $V = N^3$.

As for the exponential pre-factor $\exp(i\pi p_{\mu}/N) = \exp(i\pi p_{\mu} a/L)$,
it allows one to obtain an improved lattice Landau gauge condition \cite{Alles:1996ka},
i.e.\ the continuum condition $\partial \cdot A = 0$ is recovered with
corrections ${\cal O}(a^2)$, instead of the corrections ${\cal O}(a)$
obtained when this pre-factor is neglected.
This exponent also appears naturally in a weak-coupling expansion of
lattice gauge theory \cite{Rothe:1997kp}.
Note that this factor is a discretization correction. Indeed, it goes to
1 in the limit $a\to 0$ while keeping the physical lattice size $L=aN$ fixed.
Furthermore, it is equal to 1 in the infrared limit $p \to 0$.
This factor cancels when evaluating the scalar part of the gluon propagator
[see Eq.\ (\ref{eq:Dofp}) below],
but in general not for vertices. An exception is e.g.\ the orthogonal
configuration of the ghost-gluon vertex considered below (see Section \ref{svert}).
Numerical studies \cite{Alles:1996ka,Skullerud:1998ec} have verified
that this factor is also necessary in order to obtain the correct tensor
structure of Green's functions involving gluon fields.

After contracting Eq.\ \pref{gpbas} with a transverse projector and a unit matrix in color space,
the scalar part of the gluon propagator is given by
\be
D(p)=\frac{1}{V \, {\cal N}}\sum_{\mu,a}<\left[\Re A_\mu^a(p)\right]^2+\left[\Im A_\mu^a(p)\right]^2> ,
\label{eq:Dofp}
\ee
where $\Re A_\mu^a(p)$ and $\Im A_\mu^a(p)$ indicate, respectively,
the real and the imaginary part of $A_\mu^a(p)$
and the normalization ${\cal N}$ is given by $dN_c$ for $p>0$ and by $(d-1)N_c$ for $p=0$.
The propagator is by definition inherently positive semi-definite.
Let us recall that in minimal Landau gauge one has
\be
\sum_{\mu} P_\mu A_\mu(p)=0 ,
\label{eq:landaugauge}
\ee
where the components of the physical momenta
(denoted by capital letters when they are in lattice units) are given by
\be
P_\mu=2\sin\frac{\pi p_\mu}{N_\mu}.
\ee
Results will be presented as a function of the magnitude of the physical momentum $p = | P | / a$ (in GeV).
Note that, with our notation, the continuum gluon propagator is obtained
considering the product $\beta \, a^{2} D(k)$.
Indeed in $d$ dimensions one has $\beta = 2 N_c/(g^2 a^{4-d})$. Also, with $N_c = 2$,
the lattice quantity $2 A^a_{\mu}(x) / (g a) = \sqrt{\beta/a^{d-2}} A^a_{\mu}(x)$
goes to the continuum quantity $A^a_{\mu}(x)$ in the formal continuum limit $a \to 0$.
In the same limit, $\sqrt{\beta \, a^{d+2}} A^a_{\mu}(p)$ converges to the continuum momentum-space
gluon field $A^a_{\mu}(p)$. Thus, for any dimension $d$, the lattice quantity $\beta \, a^{2} D(k)$
converges to the continuum gluon propagator in momentum space.

It is necessary to evaluate the gluon propagator for various momentum configurations in order to
determine the vertices. In addition, some extra momentum configurations have been considered
to check effects related to the breaking of rotational invariance. The various kinematical
configurations are given in Table \ref{ppconf}. Let us note that, in general, for a given
momentum $p$ one has $A^a_\mu(p) \neq A^a_\nu(p)$, when $\mu \neq \nu$. For example, if the
momentum is aligned along the $\mu$-direction, then $A^a_\mu(p)$ vanishes for each color component
due to the relation (\ref{eq:landaugauge}) or, equivalently, due to the constancy of the
quantity $q_\mu(x_\mu)$ defined in Eq.\ (\ref{eq:defqmu}).

\begin{table}
\caption{\label{ppconf} Kinematical momentum configurations considered for the propagators.
The columns {\em Gluon} and {\em Ghost} identify whether these configurations have been used for the
evaluation of the propagators. The columns $n_x$, $n_y$ and $n_z$ give the components
of the momentum. The {\em Type} is used for later identification of classes of momenta.
The quantities $p_\mathrm{min}$ and $p_\mathrm{max}$ (in MeV) are the
smallest and the largest momenta. The smallest momentum is
evaluated at $\beta=4.2$, while the largest is evaluated at $\beta=6.0$
(in both cases we used the $30^3$ lattice volume).
The variables $n$ and $m$ run independently over all possible positive integer values $0, \ldots, N/2$
(but for the ghost propagator one cannot consider the zero momentum case).
}
\begin{ruledtabular}
\begin{tabular}{|c|c|c|c|c|c|c|c|}
  {\em Type} & $n_x$ & $n_y$ & $n_z$ & $p_\mathrm{min}$ & $p_\mathrm{max}$ & {\em Gluon} & {\em Ghost} \cr
\hline
Plane & $n$ & $m$ & 0 & 238 & 4902 & $\sqrt{}$ & $\sqrt{}$ \cr
\hline
Plane & $n$ & $-n$ & 0 & 336 & 4902 & $\sqrt{}$ & \cr
\hline
Plane & $-n$ & 0 & $n$ & 336 & 4902 & $\sqrt{}$ & $\sqrt{}$ \cr
\hline
Plane & 0 & $n$ & $-n$ & 336 & 4902 & $\sqrt{}$ & $\sqrt{}$ \cr
\hline
Diagonal & $n$ & $n$ & $n$ & 411 & 6004 & $\sqrt{}$ & $\sqrt{}$ \cr
\end{tabular}
\end{ruledtabular}
\end{table}

\begin{figure}
\includegraphics[width=\linewidth]{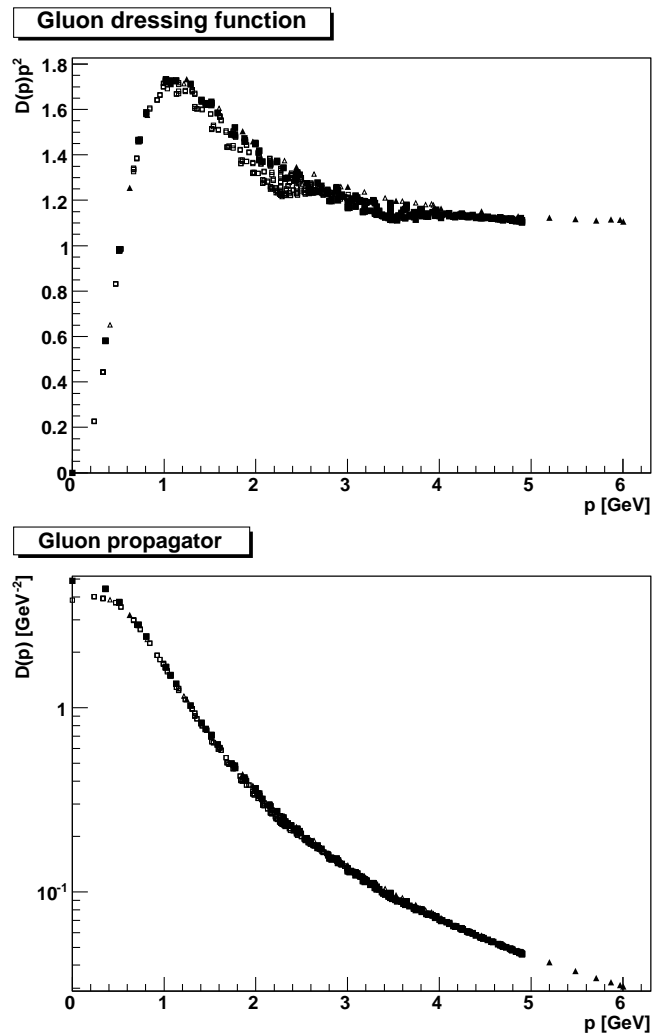}
\caption{\label{fgp}The gluon dressing function $D(p)p^2$ (top) and the gluon propagator $D(p)$
(in GeV$^{-2}$, bottom) as a function of $p$ (in GeV).
Open and full symbols correspond respectively to $\beta=4.2$ and to $\beta=6.0$.
In both cases we consider the lattice volume $V = 30^3$. Squares and
triangles correspond to plane and diagonal momentum configurations, respectively.
}
\end{figure}

The results for $D(p) p^2$ and $D(p)$ are shown in Fig.\ \ref{fgp} for the $30^3$ lattice volume
and the two $\beta$ values considered. The statistical errors are below $1\%$,
thus systematic errors are clearly visible. In particular, violation of
rotational symmetry is the dominant effect and is of the order of several percent.
In addition, finite-physical-volume effects are visible, especially in the far infrared,
when going from $\beta=4.2$ to $\beta=6.0$.
The latter effect is much more pronounced for the $20^3$ lattices.
Altogether, the systematic errors are the dominating effect at this level of statistics.

At the largest momentum (for $\beta=6.0$) the propagator is about
$6\%$ or $5.5\%$ larger than its re-summed one-loop perturbative value \cite{Maas:2004se}
(which itself deviates from the tree-level value by about $3.5\%$)
on the $20^3$ and the $30^3$ lattices, respectively.
This discrepancy cannot be explained by removing tadpole contributions
from the definition of the gluon field (see for example \cite{Bloch:2003sk}).
Indeed, if we consider \cite{Lepage:1992xa} the gauge
invariant definition of the tadpole factor given by $u_0 = <P>^{1/4}$, where $<P>$ is the
average value for the plaquette, we find (see Table \ref{conf})
that the gluon propagator is multiplied by $1/u_0^2 \approx 1.1610$ at $\beta = 4.2$ and
by $1/u_0^2 \approx 1.1011$ at $\beta = 6.0$.
This clearly enhances the discrepancy between the data and the re-summed leading-order
perturbation theory.
On the other hand, one should recall that, by changing the lattice discretization
for the gluon field (see for example \cite{Cucchieri:1999dt} and references therein),
one finds a gluon propagator that differs by a global multiplicative constant.
In particular, one can easily find definitions of the gluon field for which the
gluon propagator is sensibly smaller than that obtained with the standard definition
(see for example Table 1 in Ref.\ \cite{Cucchieri:1999ky}). Of course, different lattice
discretizations converge to a common result as the continuum limit is approached
\cite{Cucchieri:1999ky}.

Finally, we note that the combinations of lattice size and $\beta$ values considered
here are not
sufficient to reach the infrared regime, where the bending over of the propagator has been
observed \cite{Cucchieri:2003di}. Nevertheless, one clearly sees from Fig.\ \ref{fgp}
that the propagator is less singular than $1/p^2$ for momenta smaller than about 1 GeV.

\subsection{Ghost propagator}

The numerical evaluation of the ghost propagator is considerably more complicated than that of
the gluon propagator. Indeed, one has to evaluate
\be
D_G^{ab}(p)=\frac{1}{V}< (M^{-1})^{ab}(p) >,
\label{eq:DofG}
\ee
where $M^{ab}(x,y)$ is the Faddeev-Popov operator, defined in the continuum as
\be
-\pdm D^{ab}_\mu=\delta(x-y) (-\pd^2\delta^{ab}+gf^{abc}\pd_\mu A_\mu^c) \, .
\ee
Here $f^{abc}$ are the structure constants of the
$SU(N_c)$ gauge group. In Landau gauge, on the lattice, this operator is a matrix
(with color and space-time indices), defined
by its action on a scalar function $\omega^b(x)$ (with color index $b$) as
\cite{Zwanziger:1993dh}
\bea
\! M^{ab}(x,y)\omega^b(y)\!\!&=&\!\!\delta_{xy}\sum_{\mu}\{G_\mu^{ab}(y)[\omega^b(y)-\omega^b(y+e_\mu)]\nn\\
&& \quad -G_\mu^{ab}(y-e_\mu)[\omega^b(y-e_\mu)-\omega^b(y)]\nn\\[2mm]
&& \quad +\sum_{c}f^{abc}[A_\mu^b(y)\omega^c(y+e_\mu) \nn\\
&& \quad \;\; -A_\mu^b(y-e_\mu)\omega^c(y-e_\mu)]\} .
\eea
Here, the sum over repeated indices ($y$ and $b$) is understood and $G_\mu^{ab}(x)$ is given by
\be
G^{ab}_\mu(x)=\frac{1}{8}\tr\left(\{\sigma_a,\sigma_b\}\left[U_\mu(x)+U_\mu(x)^+\right]\right),
\ee
i.e.\ it is proportional to $\delta^{ab}$.
Let us recall that in minimal Landau gauge \cite{Zwanziger:1993dh}
the gauge-fixed configurations are
inside the first Gribov horizon, implying that the Faddeev-Popov operator $M^{ab}(x,y)$
is symmetric
and positive (in the subspace orthogonal to the trivial and constant zero modes).

In order to evaluate the Fourier transform of the inverse operator \footnote{In the general
case one considers $p\neq q$. However, in the case of the ghost propagator, one has
$p=-q$ due to momentum conservation. Note that, again, we do not divide by the lattice
volume $V$ when considering the Fourier transform.}
\be
(M^{-1})^{ab}(p,q)=\sum_{x,y} e^{2\pi i (px+qy)/N}(M^{-1})^{ab}(x,y)
\ee
the matrix inversion has been performed using the point source $\delta^{ac}(\delta_{x0}-1/V)$
(see Ref.\ \cite{Boucaud:2005gg}). Compared to the inversion using a plane-wave source \cite{Cucchieri:1997dx},
this method has the advantage of only $N_c^2-1$ inversions per configuration, independently of the
number of momenta, instead
of $N_c^2-1$ inversions per configuration and for each momentum considered. However, as we will
see below, statistical fluctuations are significantly enhanced at large momenta \cite{Boucaud:2005gg}.
Also, note that this procedure is ambiguous
with respect to the sign of the resulting propagator \footnote{When considering the plane-wave source the
ghost propagator is the expectation value of a positive operator on the plane-wave state, i.e.\
the result is always positive-definite. This is not true with the point-source method. Indeed, one can
obtain (on a given lattice configuration) a negative value for the quantity $(M^{-1})^{ab}(p)$.
Nevertheless, on average, the ghost propagator $D_G(p)$ is positive for all values of the momentum $p$
also when using the point-source method.}
(or of the eigenvalues). Thus, the sign has to be assigned by hand.

Since the Faddeev-Popov operator $M^{ab}(x,y)$ is symmetric and positive,
the matrix inversion can be performed using a conjugate-gradient (CG)
method (see for example \cite{Meurant:2006}).
From the numerical point of view, there
are two issues to be careful about.
First of all, as finite-precision arithmetic is used, it is possible for the solution
to develop a component along the (constant) zero modes of the Faddeev-Popov operator.
Thus, it is necessary to reorthogonalize the solution at each CG iteration with respect
to this subspace.

The second aspect is related to the convergence of the CG method. Indeed, in finite precision arithmetic,
the CG method normally loses orthogonality to the Krylov subspace already spanned. Thus, the magnitude
of the residual can be incorrect. To compensate for this, an additional, albeit technical,
quality criterion can be considered (see \cite{Meurant:2006} section 5.3 for details).
As a convergence test, we check if the average of the components of the residual is less than $10^{-12}$
and if the additional quality parameter \footnote{In the present case,
due to memory restrictions, we only consider subsequences of the CG iteration of length 1.}
is less than $10^{-13}$ between two CG iterations.
This is usually achieved after several tens to a few hundreds of iterations.
We have also checked that much
weaker convergence tests do not change the propagators by more than a few percent.
Let us note that one CG inversion of the Faddeev-Popov matrix is much faster than the gauge-fixing
procedure when considering large lattice volumes.

\begin{figure}
\includegraphics[width=\linewidth]{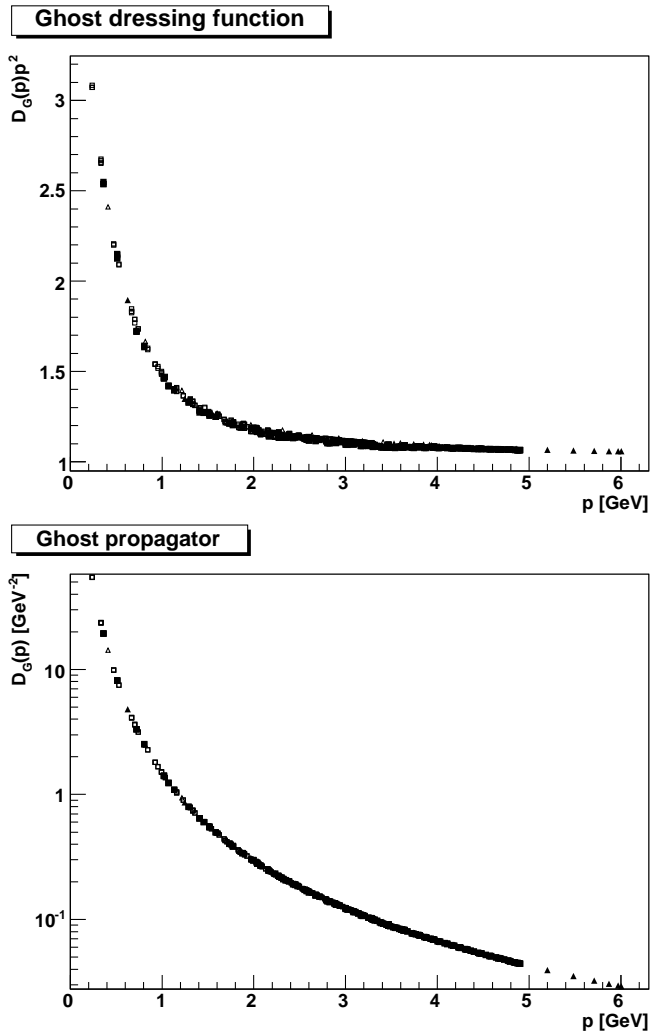}
\caption{\label{fghp} The color-averaged diagonal part of the ghost dressing function $D_G(p)p^2$
(top) and of the ghost propagator $D_G(p)$ (in GeV$^{-2}$, bottom) as a function of the momentum $p$ (in GeV).
Open and full symbols are used for $\beta=4.2$ and $\beta=6.0$, respectively.
In both cases we consider the lattice volume $V = 30^3$. Squares
and triangles correspond to plane and diagonal momentum configurations, respectively.
}
\end{figure}

The results for the color-averaged diagonal ghost propagator are shown in Fig.\ \ref{fghp} for the
lattice volume $V=30^3$.
Again, the statistical error is essentially negligible compared to the systematic effects.
Note that the consequences of violation of rotational symmetry are much less severe than
in the gluon case, although they are still the dominant source of systematic errors.
As the ghost is a scalar, this was somehow expected. Also, the finite-physical-volume errors
are somewhat smaller than in the gluon case. Comparing the lattice data to the perturbative
predictions \cite{Maas:2004se} at the largest momentum (for $\beta=6.0$) we find that
the former are larger by about $6\%$ and $5.5\%$ for the
lattice volumes $V=20^3$ and $V=30^3$, respectively.
In this case, perturbation theory only
deviates about $1\%$ from the asymptotic tree-level value at this momentum.
If one considers tadpole improvement \cite{Bloch:2003sk}, then the ghost propagator gets multiplied
by $u_0$, i.e.\ by 0.92807 at $\beta = 4.2$ and by 0.95298 at $\beta = 6.0$, leading
to a better agreement with perturbation theory.

Here we didn't try to fit the data for the ghost propagator and obtain an estimate for its
infrared exponent. Nevertheless, from Fig.\ \ref{fghp} one clearly sees that
$D_G(p)$ is more singular than $1/p^2$.

\begin{figure}
\includegraphics[width=\linewidth]{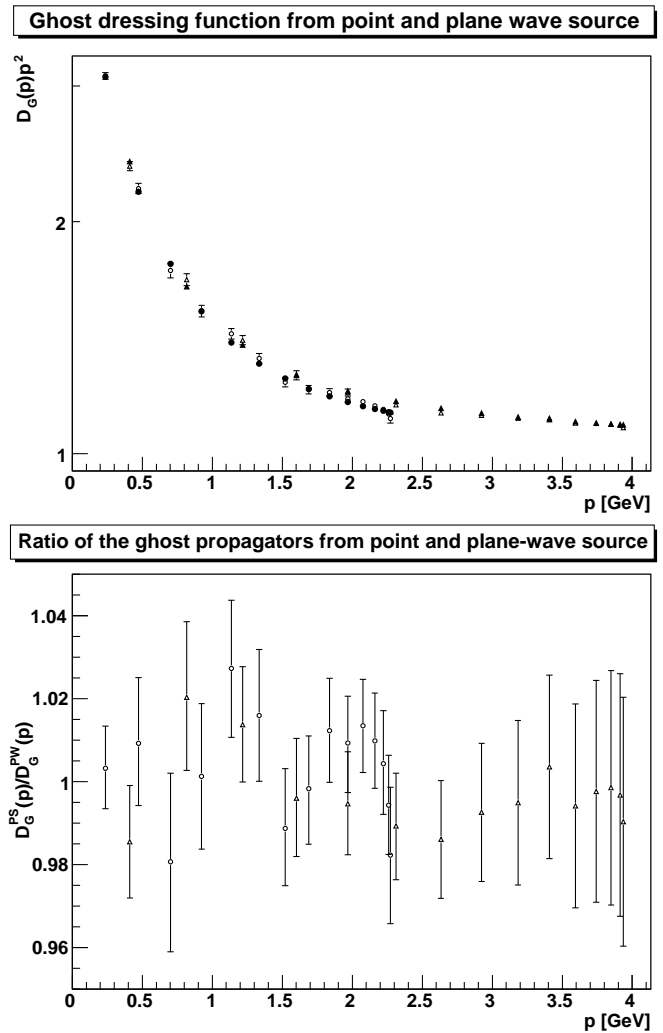}
\caption{\label{fmethod-comp} In the top panel we show the ghost dressing function 
$D_G(p)p^2$ obtained using the plane-wave source (full symbols) and the point source
(empty symbols) for $V=30^3$ and $\beta=4.2$. Note the logarithmic scale on the
$y$ axis. In the bottom panel we report the
ratio of the point-source data (PS) with the the plane-wave-source results (PW)
for the ghost propagator. In both panels, circles denote momenta of type $(0,0,p)$, triangles
indicate diagonal momenta and the quantities are considered as a function of the momentum
$p$ (in GeV). Here, for both methods, we considered 380 configurations.
}
\end{figure}

We also compared the results obtained using the two different inversion methods
mentioned above (i.e.\ point-source and plane-wave-source methods). In the
second case the inversion has been done using a CG method with even-odd preconditioning
and a convergence test given by $| r |^2 / | r_0 |^2 \leq 10^{-8}$, where $r$ and $r_0$
are the final and the initial residuals, respectively. The use of the even-odd preconditioning
usually reduces the number of CG iterations by about a factor 2.
We find that the results from these two methods agree on the average for the ghost
propagator as well as for the ghost-gluon vertex, discussed below.
As can be seen in Fig.\ \ref{fmethod-comp}, the ghost propagator obtained using the
point source oscillates around the (smoother) result obtained using the plane-wave source.
This oscillatory behavior is stronger for the ghost-gluon vertex, but vanishes in both cases when
the statistics is increased, albeit much more slowly for the latter case.

\begin{figure}
\includegraphics[width=\linewidth]{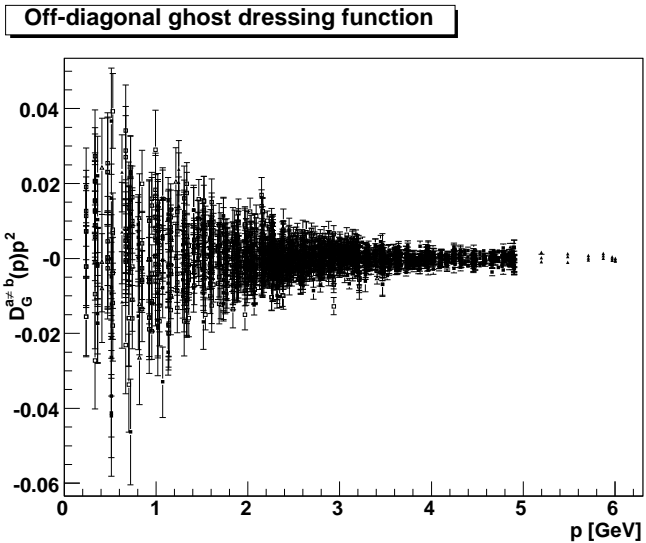}
\caption{\label{foffghp}The real part of the color upper-triangular matrix elements
of the ghost dressing function $D_G^{ab}(p)p^2$ as a function of the momentum $p$ (in GeV).
Open and full symbols indicate $\beta=4.2$ and $\beta=6.0$, respectively.
In both cases we consider the lattice volume $V = 30^3$.
Squares and triangles correspond to plane and diagonal momentum configurations, respectively.
}
\end{figure}

\begin{figure}
\includegraphics[width=\linewidth]{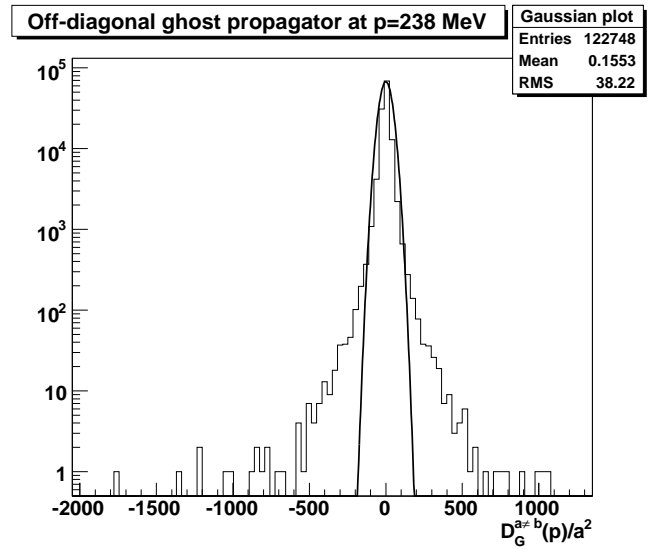}
\caption{\label{fng} A histogram of the real part of the
off-diagonal elements of the ghost propagator (in lattice units)
evaluated at the smallest momentum for the lattice volume $V=30^3$ at $\beta=4.2$.
We also plot a Gaussian with the same mean and standard deviation as the histogram of the data.
}
\end{figure}

An important assumption in functional calculations is that the ghost
propagator is color diagonal \footnote{We also checked that the real part
of the off-diagonal components of the
gluon propagator is essentially zero within statistical errors and that its
central value decreases with increasing statistics.
Of course, the imaginary part vanishes identically.}.
In Fig.\ \ref{foffghp} we show
results for the real part of the off-diagonal
components of the ghost propagator \footnote{Let us recall that the
fluctuations of the {\em imaginary} part of the
off-diagonal elements of the ghost propagator are
connected to the possible existence of a ghost condensate \cite{Cucchieri:2005yr}. In
particular, non-Gaussian fluctuations could indicate the existence of a spontaneous-symmetry
breaking and a non-vanishing value for this condensate. (For a different
interpretation, see Ref.\ \cite{Furui:2006rx}.) In this work we do not present data for
the ghost condensate.}.
The majority of the points are compatible with a null result, within the $67\%$-confidence interval,
and the mean value of the (real part of the)
off-diagonal propagator decreases for all momenta with increasing statistics.
On the other hand, these fluctuations exhibit significantly enhanced tails
(see Fig.\ \ref{fng}). We checked that these large fluctuations are in most cases
related to large values of the color-diagonal part of the ghost propagator and to configurations for which
gauge fixing to Landau gauge required many more iterations than in the average case.
Thus, these tails could be related to the exceptional configurations observed in \cite{Sternbeck:2005tk}.
Let us note that, by using the point-source method \cite{Boucaud:2005gg} for the inversion,
one needs very large statistics in order to see a null average for the off-diagonal points
in the infrared.
Also, using this method, the real (respectively, imaginary) part of the inverse
Faddeev-Popov matrix $(M^{-1})^{ab}(p,-p)$ is symmetric (respectively anti-symmetric)
only on average and not for each lattice configuration, as is the case
when using the plane-wave source \cite{Cucchieri:2005yr}.

\subsection{Eigenvalue spectrum of the Faddeev-Popov operator}

The CG method has a close relationship
with the so-called Lanczos algorithm, which can be used to extract
the eigenvalues of a matrix \cite{Meurant:2006}.
As a consequence, it is possible to determine
the ghost propagator and simultaneously obtain information on the eigenvalue
spectrum. Note that only the {\em exact} CG algorithm
is guaranteed to obtain the correct spectrum (up to degeneracy)
of the Faddeev-Popov matrix after at most $(N_c^2-1)V$ inversion steps.
On the other hand, working in finite-precision arithmetic (see Section 4 in \cite{Meurant:2006}),
the algorithm typically does not deliver a sufficiently
accurate numerical approximation to all eigenvalues of this matrix.
Indeed, as the number of iterations increases, the method has the property that
the extremal eigenvalues evaluated are progressively improved
approximations of the extremal eigenvalues of the matrix considered \cite{Parlett}.
At the same time, since the procedure is stopped before all eigenvalues are found,
the middle of the spectrum is usually not reliable and under-populated.
Moreover, each eigenvalue will be found only once, independently
of its degeneracy.
Still, when averaging over many configurations --- i.e.\
when considering a histogram of the eigenvalues
found for all the configurations, normalized by the total
number of eigenvalues \footnote{Doing this normalization for each inversion
and each configuration separately does not yield a different result
within the statistical errors.} --- one should be able to obtain an idea
of the density of the eigenvalues at both ends of the spectrum.
In particular, one can verify whether the shape of the eigenvalue density
(at small eigenvalues) is flat or increases with the magnitude of
the eigenvalues $\omega$, thus gaining information on the validity
of the Gribov-Zwanziger scenario. On the other hand, the upper end
of the spectrum, which is also obtained, is essentially determined
by perturbative contributions and thus not as interesting.

\begin{figure}[ht]
\includegraphics[width=\linewidth]{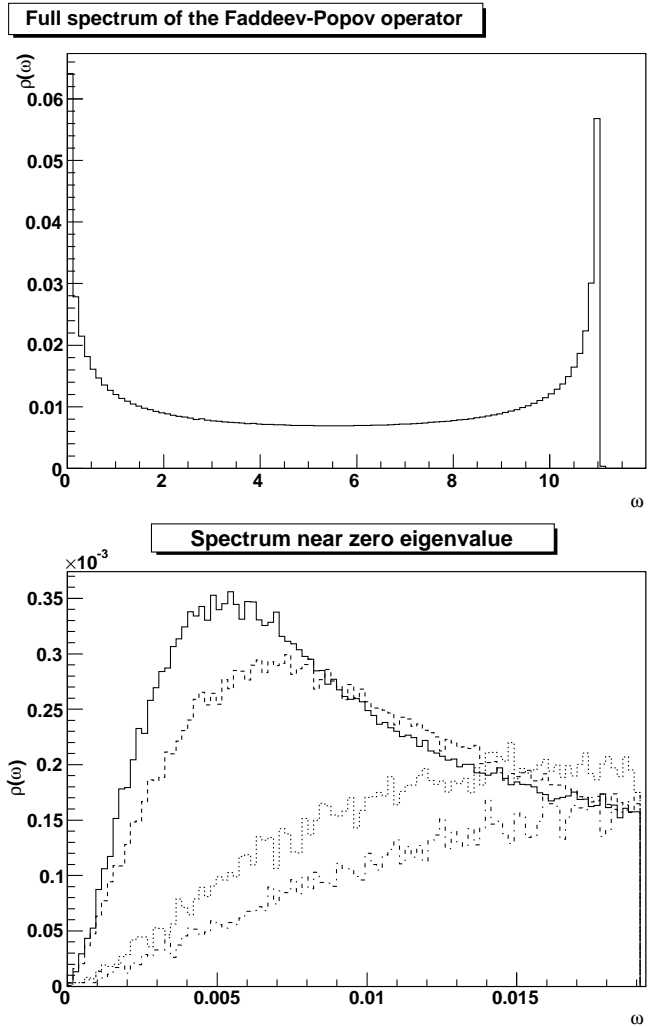}
\caption{\label{feigen} The density $\rho(\omega)$ of the eigenvalues of the Faddeev-Popov operator
(using the Lanczos method) as a function of their magnitude $\omega$ (in lattice units),
normalized by the total number of eigenvalues.
In the top panel we show
the complete results while in the bottom panel only the density for small eigenvalues is presented.
In both cases we considered 100 bins (this explains the different scale on the y axis).
The solid line (in both panels) refers
to the lattice volume $V=30^3$ at $\beta=4.2$. In the bottom panel, the dashed line correspond to
$V=30^3$ at $\beta=6.0$, the dotted line to $V=20^3$ at $\beta=4.2$ and the dashed-dotted
line to $V=20^3$ at $\beta=6.0$. The total number of eigenvalues determined are
4374822 ($V=20^3$, $\beta=4.2$), 11312874 ($V=30^3$, $\beta=4.2$), 
3549202 ($V=20^3$, $\beta=6.0$) and 9862721 ($V=30^3$, $\beta=6.0$).
}
\end{figure}

The results are shown in Fig.\ \ref{feigen} (top panel). The full spectrum is shown for completeness.
As said above, the
small density of the spectrum for intermediate magnitude $\omega$ of the eigenvalues is very likely
an artifact of the algorithm.
For very small magnitude $\omega$
(see Fig.\ \ref{feigen}, bottom panel) the spectrum shows a linear increase with $\omega$.
The steepness is
much larger on the larger lattices and it also seems to increase with the physical volume, i.e.\
as $\beta$ decreases.
These results are in qualitative agreement with those reported in Ref.\ \cite{Sternbeck:2005vs}
for the four-dimensional case. Note, however, that a linear increase at small momenta is
different from the result obtained in Coulomb gauge in four dimensions \cite{Greensite:2004ur},
where a power-like behavior has been observed.

\begin{table}
\caption{\label{tab:eigen} Largest and smallest eigenvalues of the Faddeev-Popov matrix
(respectively, $\omega_l$ and $\omega_s$, both in lattice units) for each lattice
volume $V$ and coupling $\beta$. We also report the corresponding eigenvalues of
the lattice Laplacian (for the same lattice side $N$). For the lattice volume
$V = 20^3$ we consider 200 configurations at $\beta = 4.2$ and 400 at $\beta = 6.0$;
for $V = 30^3$ we have 380 and 500 configurations at $\beta = 4.2$ and $6.0$, respectively.
}
\begin{ruledtabular}
\begin{tabular}{|c|c|c|c|c|c|}
Volume & $\beta$ & $\omega_l$ & $\omega_s$ & $\omega_{l, Lapl.}$ & $\omega_{s, Lapl.}$ \\  
\hline
$20^3$ & 4.2 & 11.1218(9) & 0.0072(2) & 11.9261 & 0.0246 \\
\hline
$30^3$ & 4.2 & 11.1337(5) & 0.00275(6) & 11.9671 & 0.0110 \\
\hline
$20^3$ & 6.0 & 11.4050(4) & 0.0102(2) & 11.9261 & 0.0246 \\
\hline
$30^3$ & 6.0 & 11.4149(2) & 0.00353(7) & 11.9671 & 0.0110 \\
\end{tabular}
\end{ruledtabular}
\end{table}

\begin{figure}
\includegraphics[width=\linewidth]{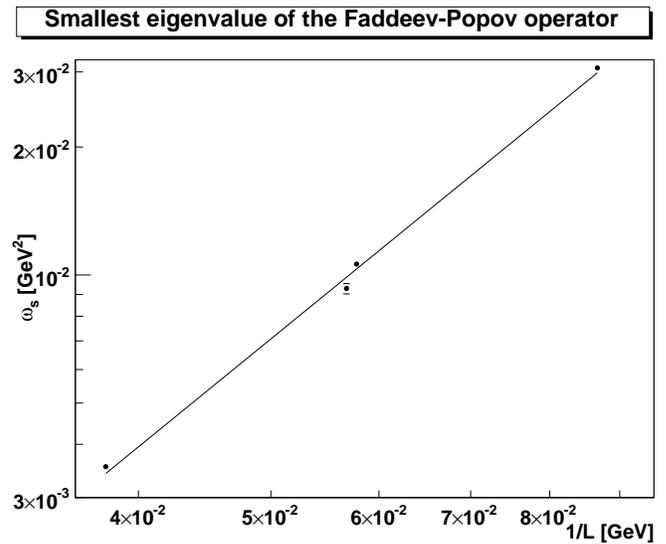}
\caption{\label{fig:eigen} Plot of the smallest eigenvalue $\omega_s$ (in GeV$^2$)
of the Faddeev-Popov matrix, as a function of the inverse lattice side $1/L$
(in GeV), and of the fitting function $a / L^c$ with $a = 18.127$ and $c = 2.6152$.
}
\end{figure}

Using a preconditioned CG method we have also evaluated the smallest and the largest
eigenvalues of the Faddeev-Popov matrix $M$. Results are reported in Table \ref{tab:eigen}.
For comparison we also give the smallest and the largest eigenvalues of the lattice
Laplacian $- \Delta$ for the same lattice size. Let us recall that in $d$ dimensions,
the smallest eigenvalue of $-\Delta$ is given by $4 \sin^2(\pi / 2 N)$, while the
largest is equal to $ 4 d \sin^2[\pi (N-1) / 2 N]$. We see that the largest eigenvalue
of $M$ depends weakly on the lattice volume $V$ and on the the coupling $\beta$.
On the other hand, for a given $\beta$, the smallest eigenvalue decreases faster
than the corresponding eigenvalue of $- \Delta$, as the lattice side increases.
This is in agreement with the results reported in \cite{Cucchieri:1997dx}.
Also, if we consider the smallest eigenvalue of $M$ in physical units, i.e.\
if we multiply it by $1/a^2$, we find the results reported in Fig.\ \ref{fig:eigen}.
If one tries a fit to data using the Ansatz $a / L^c$ we obtain that
the data are well fitted for $c = 2.62(8)$ (with $\chi /d.o.f. = 1.4$). This seems
to suggest that for these lattice volumes and $\beta$ values we are in the scaling
region for $\omega_s$ and that this eigenvalue goes to zero when the infinite-volume
limit is approached. As a consequence, in the continuum limit, the average lattice
Landau configuration should belong to the first Gribov horizon, supporting the
Gribov-Zwanziger mechanism of confinement.

\section{Vertices}\label{svert}

In Yang-Mills theory (without valence and sea quarks)
there are two non-vanishing 3-point Green's functions:
the three-gluon vertex and the ghost-gluon vertex. Compared to the propagators, these
are quite complicated since there are now three kinematical variables.
In addition, for the three-gluon vertex there exist a large number of tensor structures
\cite{Ball:1980ax}. Therefore, it is impractical to investigate all tensor structures
for all possible kinematical configurations.
Furthermore, due to kinematics, the interesting signal --- i.e.\ an appropriately
defined dimensionless function --- is usually suppressed at large momenta as
$1/p^6$ instead of as $1/p^2$ (except for some particular momentum configurations 
for which it is only suppressed
by $1/p^4$). Thus, it is in general quite difficult to obtain a good signal/noise ratio
at large $p$.

Let us recall that, due to momentum conservation, the three momenta of the vertices
always lie in one plane and are completely characterized by the size of two of the
momenta and by the angle $\phi$ between them.
Two particular momentum configurations are of special importance.
In the first case two of the momenta are orthogonal with respect to each other
(this configuration will be called here the {\em orthogonal} one).  This
configuration enters into loop integrals with maximum angular weight \footnote{Let us recall
that this angular weight is proportional to $\sin(\phi)$ in 3d.}
and, as we will see below, has the advantage of relatively small statistical fluctuations.
In this case we take the two orthogonal momenta respectively along the $x$ and $y$ axis,
considering all possible magnitudes of both momenta independently.
If one of these two momenta vanishes, then one recovers the kinematical configuration
used in several previous lattice studies of these vertices
\cite{Parrinello:1994wd,Alles:1996ka,Cucchieri:2004sq,Boucaud:2000ey}. 

In the second configuration (called here the {\em equal} one) the three momenta have the
same magnitude and the angle $\phi$
is equal to $\pi/3$. The vertex is then a function of only one variable.
For this configuration the infrared behavior of the $n$-point Green's
functions can be taken to be the limit where all $n$ momenta vanish in the same way, i.e.\
only one scale has to be considered. In this case there are predictions for the behavior of
the $n$-point Green's functions in the infrared limit, from studies using Dyson-Schwinger equations
(DSEs) \cite{Alkofer:2004it,Schleifenbaum:2006bq}. Of course, on the lattice, it is not possible
(in general) to select three equal momenta,
for example, in the $x-y$ plane. However, one can consider the
momenta $(p,-p,0)$, $(-p,0,p)$ and $(0,p,-p)$, which all have the (lattice) size $\sqrt{2} P$
with $P = 2\sin(\pi p/N)$.

Let us recall that, due to the invariance of the lattice theory under hyper-cubic transformations,
the results should be invariant by reflection of the momenta.
This implies that --- with the exception of $p=0$ and of the mid-point
on even lattices --- we can average over two different kinematical configurations
for each physical momentum.
For the vertices, due to momentum conservation, we
can apply the reflection transformation only to the independent momenta. In particular,
in the orthogonal case, we can apply the reflection independently on the two orthogonal momenta,
i.e.\ --- for each physical momentum configuration ---
we can average over four different kinematical configurations.
On the contrary, in the equal-momenta case, the reflection has to be applied at the same
time on the three momenta, i.e.\ each time we can average over two
different kinematical configurations. Let us note that this averaging
allows us to cancel exactly, on each lattice configuration,
contributions that would otherwise be zero only on average, yielding purely
imaginary vertices.

Finally, one has to consider the tensor structure of the vertices, which is not straightforward
for the three-gluon vertex. In functional-method studies it is particular interesting to
know how different a vertex is from its tree-level structure. Thus, we consider here
the projection of the vertices on their tree-level values, obtaining a scalar function
normalized to 1 when the tree-level vertex and the full vertex are equal.

\begin{widetext}

\vskip 2mm

\subsection{Three-gluon vertex}

The three-gluon vertex $\Gamma_{\mu\nu\rho}^{\tl\indexsep A^3\indexsep abc}(p,q,k)$ (with $k=-p-q$)
at tree-level in the continuum has the form
\be
\Gamma_{\mu\nu\rho}^{\tl\indexsep A^3\indexsep abc}(p,q,k)\,=\,-igf^{abc}[(q-k)_\mu\delta_{\nu\rho}
+ (k-p)_\nu\delta_{\mu\rho} + (p-q)_\rho\delta_{\mu\nu}] \, ,
\label{eq:gammatree}
\ee
while on the lattice one finds \cite{Rothe:1997kp}
\be
\Gamma_{\mu\nu\rho}^{\tl\indexsep L\indexsep A^3\indexsep abc}(p,q,k)\,=\,-igf^{abc}
\, e^{i\pi (p_{\mu} + q_{\nu} + k_{\rho})/N} \,[(\widetilde{q-k})_\mu
        \delta_{\nu\rho}\cos(\hat{p}_\nu) + (\widetilde{k-p})_\nu\delta_{\mu\rho}\cos(\hat{q}_\rho)
   + (\widetilde{p-q})_\rho\delta_{\mu\nu}\cos(\hat{k}_\mu)] \, .
\label{eq:gammarothe}
\ee
Note that the cosine factors in the above equation go to 1 in the formal
continuum limit $a \to 0$, for a fixed physical lattice side $L = N a$. 
The same applies to the exponential pre-factor.
Here we used the notation
\bea
\tilde{p}_\mu&=&2\sin\left(\hat{p}_\mu\right)\equiv  P_\mu \\
\hat{p}_\mu&=&\frac{\pi p_\mu}{N}
\eea
with $p_\mu$ taking values $0\, , \ldots , \, N/2 \, $.
Clearly, the three-gluon vertex is totally symmetric under the simultaneous exchange of
color index, Lorentz index and momentum, and thus completely Bose-symmetric.

The full three-gluon vertex $\Gamma_{\mu\nu\rho}^{A^3\indexsep abc}$ is not directly
available on the lattice, but one can evaluate the corresponding full Green's
function \cite{Parrinello:1994wd}
\be
G_{\mu\nu\rho}^{A^3\indexsep abc}(p,q,k)=\frac{1}{V}<A_\mu^a(p)A_\nu^b(q)A_\rho^c(k)>\label{gfg3v}.
\ee
As said above, momentum conservation requires $k=-p-q$. Due to the vanishing of any vector condensates
in Yang-Mills theory, this Green's function equals the connected Green's function,
but it is still necessary to amputate it. The relation of the full vertex
$\Gamma^{A^3\indexsep def}_{\lambda\sigma\omega}(p,q,k)$ with the Green's function is then given by
\be
G_{\mu\nu\rho}^{A^3\indexsep abc}(p,q,k)=D_{\mu\lambda}^{ad}(p)D_{\nu\sigma}^{be}(q)D_{\rho\omega}^{cf}(k)
\Gamma^{A^3\indexsep def}_{\lambda\sigma\omega}(p,q,k) .
\label{eq:GofGamma}
\ee
In Landau gauge, one can extract only the transverse part of the full vertex. In order to
project the quantity above on the tree-level vertex, the following function will be evaluated
\bea
G^{A^3}(p,q,\phi)&=&\frac{\Gamma_{\mu\nu\rho}^{\tl\indexsep L\indexsep A^3\indexsep abc}(p,q,k) \;
        G_{\mu\nu\rho}^{A^3\indexsep abc}(p,q,k)}{\Gamma_{\mu\nu\rho}^{\tl\indexsep
                          L\indexsep A^3\indexsep abc}(p,q,k)
      \; D_{\mu\lambda}^{ad}(p) \; D_{\nu\sigma}^{be}(q) \; D_{\rho\omega}^{cf}(k) \; 
        \Gamma^{\tl\indexsep L\indexsep A^3\indexsep def}_{\lambda\sigma\omega}(p,q,k)}
      \label{eq:G3AAA}  \\[2mm]
&=&\frac{\Gamma_{\mu\nu\rho}^{\tl\indexsep L\indexsep A^3\indexsep abc}(p,q,k) \;
         G_{\mu\nu\rho}^{A^3\indexsep abc}(p,q,k)}{\Gamma_{\mu\nu\rho}^{\tl\indexsep
                    L\indexsep A^3\indexsep abc}(p,q,k) \;
        P_{\mu\lambda}(p) \; P_{\nu\sigma}(q) \; P_{\rho\omega}(k) \; 
        \Gamma^{\tl\indexsep L\indexsep A^3\indexsep abc}_{\lambda\sigma\omega}(p,q,k)
                  \; D(p) \; D(q) \;D(k)}\, ,\label{g3v}
\eea
\end{widetext}
where the $D$'s are the scalar gluon propagators.
Clearly, this function is equal to 1 if the full vertex is equal to the bare vertex.
Note that, by contracting the Green's function with the lattice version of the tree-level vertex
instead of the continuum version corrects for (relevant) discretization effects.
Also note that the exponential pre-factor in Eq.\ (\ref{eq:Aofp}) implies a pre-factor
$\exp[- i\pi (p_{\mu} + p_{\lambda})/N]$ for the gluon propagator $D_{\mu\lambda}^{ad}(p)$.
Thus, considering Eq.\ \pref{eq:gammarothe}, all these pre-factors cancel each other
both in the numerator and in the denominator of Eq.\ (\ref{eq:G3AAA}) above.

The normalization factor in the denominator is in some kinematical cases quite simple, but in general
very lengthy and will not be given here explicitly. Also, this normalization factor vanishes for the
largest momentum in each momentum configuration due to the cosine factors.
Thus, this momentum cannot be considered. The same applies to the case where all momenta vanish.

Let us stress that, although the quantity in Eq.\ \pref{g3v} is 1 if the full and the tree-level vertex coincide,
it will in general have contributions from tensor structures not appearing at the
tree level. The situation would be even more
complicated in the case of a vertex not totally antisymmetric in color space.
In the continuum, the general tensor structure of the total color-antisymmetric
part of the full three-gluon vertex is given by \cite{Ball:1980ax}
\begin{widetext}
\bea
\Gamma^{A^3\indexsep abc}_{\mu\nu\rho}(p,q,k)&=&-if^{abc}(A(p,q,k)\delta_\mn(p_\rho-q_\rho)
       +B(p,q,k)\delta_\mn(p_\rho+q_\rho)+C(p,q,k)(p_\nu q_\mu-\delta_\mn p \cdot q) (p_\rho-q_\rho)\nn\\
  &&+\frac{S(p,q,k)}{3} (p_\rho q_\mu k_\nu+p_\nu q_\rho k_\mu))+\mathrm{cyc.~perm.} \; ,
\eea
where $A$, $B$, $C$ and $S$ are scalar functions. (See again Ref.\ \cite{Ball:1980ax} for the
symmetry properties of these functions.) Then, one can verify that, in the continuum,
the quantity \pref{g3v} contains only contributions from the functions $A$ and $C$.
For example, in the orthogonal configuration, using the above equation one finds
\bea
G^{A^3c}\left(p,q,\frac{\pi}{2}\right)&=&\{ xyA(p,q,k)+(x+y)[2yA(k,p,q)+2xA(q,k,p) \nn\\[2mm]
&&\qquad +(xy+2y^2)C(k,p,q)+(2x^2+xy)C(k,q,p)+xyC(q,p,k)]\} / (2x^2+5xy+2y^2) \, ,
\eea
\end{widetext}
where $x=q^2$ and $y=p^2$. The expression is lengthier for other kinematical configurations.

It then only remains to determine the 3-point Green's function.
This can be done in a straightforward way using the definition \pref{gfg3v} and the
Fourier-transformed gauge fields defined in Eq.\ (\ref{eq:Aofp}).
The only problem is that the functions $A_\mu^a(p)$ vanish on average. This induces
very large fluctuations in the calculation of the Green's function,
particularly at large momenta. This problem makes the extraction of the
equal-momentum configuration quite complicated. In the orthogonal configuration
the situation is slightly better, since some field components
identically vanish even for non-vanishing momenta, as discussed in Section \ref{sprop},
reducing the statistical noise.
Also note that, in the evaluation of Eq.\ (\ref{g3v}), the Green's function and the propagators
are calculated independently and divided after averaging over
all lattice configurations. The error can then be determined by error propagation,
but due to the smallness of the statistical errors on the propagators, their errors are neglected here.

\begin{figure}
\includegraphics[width=\linewidth]{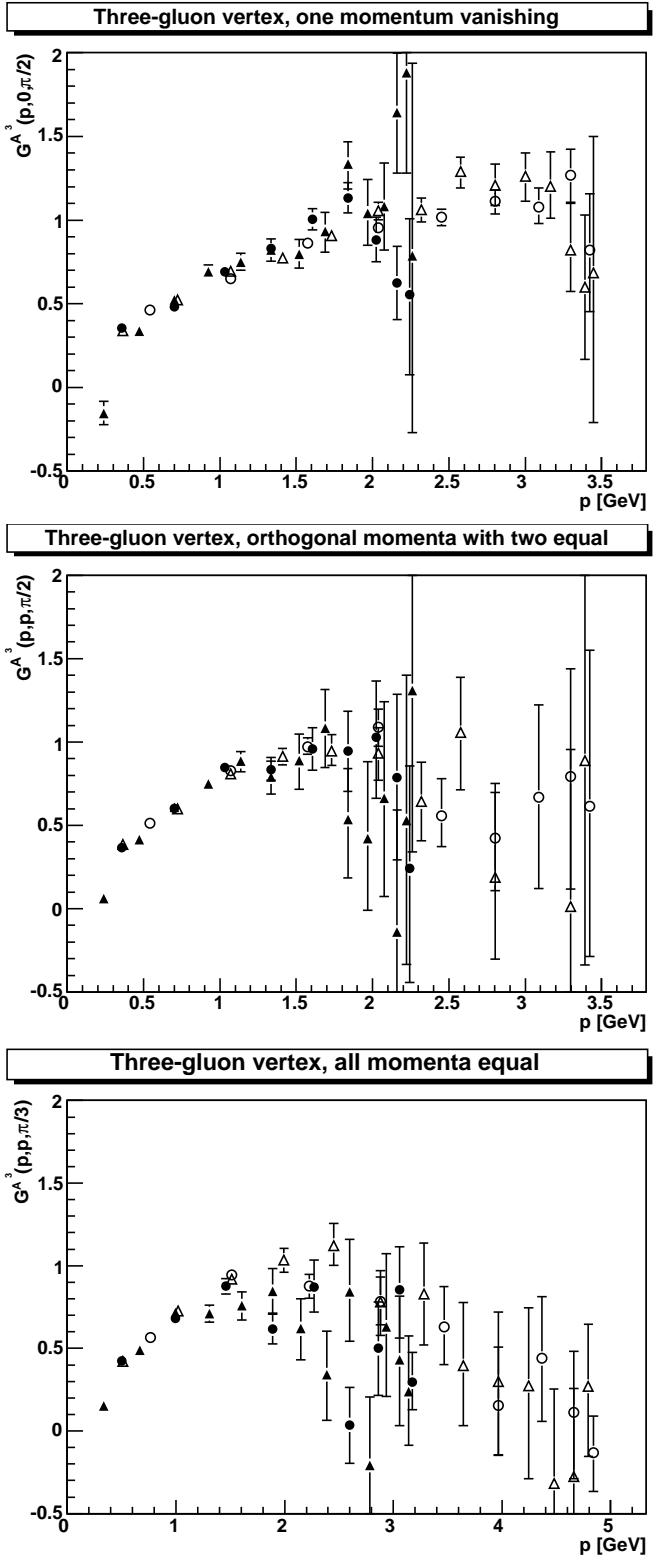}
\caption{\label{fg3v} The scalar function $G^{A^3c}(p,q,\phi)$ defined in Eq.\ (\ref{eq:G3AAA}).
Full symbols correspond to $\beta=4.2$ and open symbols to $\beta=6.0$;
circles are used for $V=20^3$ and triangles for $V=30^3$.
In the top panel we show results for the orthogonal configuration with one momentum ($q$) vanishing.
In the middle panel we consider an orthogonal configuration with
two momenta having the same magnitude ($p=q$).
In the bottom panel we plot data for the case with the three momenta equal
($p=q=k$). 
}
\end{figure}

\begin{figure}[ht]
\includegraphics[width=\linewidth]{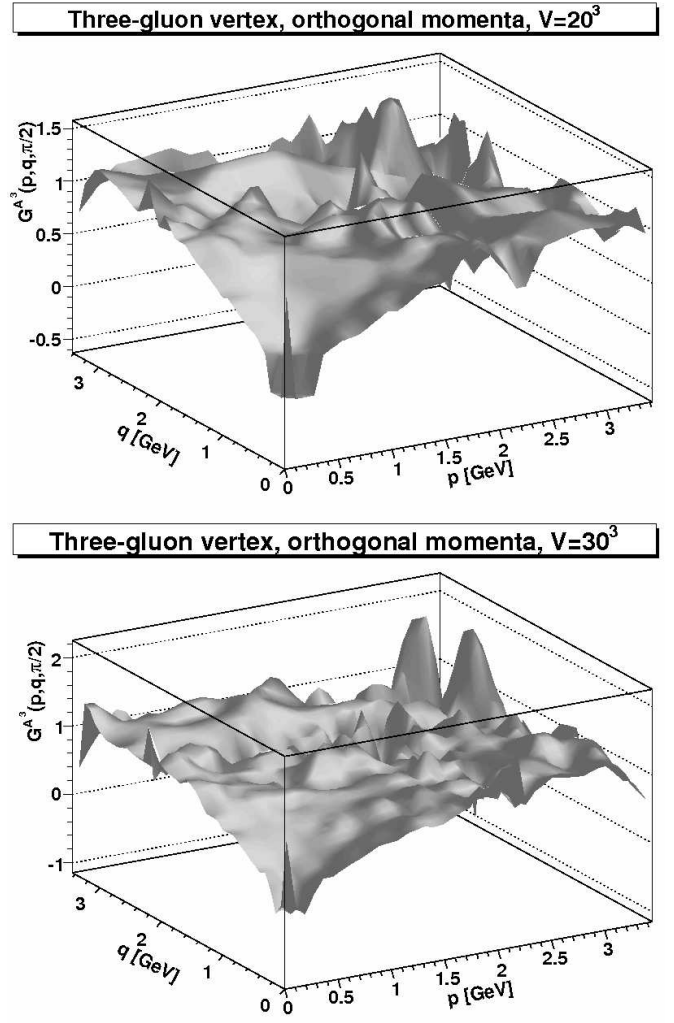}
\caption{\label{fg3vo} The scalar function $G^{A^3c}(p,q,\phi)$ defined in Eq.\ (\ref{eq:G3AAA})
as a function of the magnitude of the gluon momenta $p$ and $q$ (for the orthogonal configuration).
Here, for each data point, we plot only the central value.
The data for $\beta=4.2$ and $\beta=6.0$ are plotted together
and they are interpolated by Gouraud shading, as implemented in the ROOT package \cite{Brun:1997pa}.
In the top figure we used data for the lattice volume $V=20^3$; in the bottom one
we consider the lattice volume $V=30^3$.
Spikes indicate positions where due to fluctuations the value is outside the drawing range.
The spike at $p=q=0$ is an artifact since this quantity cannot be evaluated when the
three momenta are null.
}
\end{figure}

Let us note that, from the discussion in Section \ref{gluonprop},
we found that (with our notation) the continuum momentum-space gluon field has mass
dimension $-1-d/2$ while the gluon propagator, again in momentum space, has mass
dimension $-2$. This implies that the full Green's function defined in Eq.\ \pref{gfg3v}
has mass dimension $-3-d/2$. The same result can be obtained by considering Eq.\ 
\pref{eq:GofGamma} after observing that the three-gluon vertex 
$\Gamma^{A^3\indexsep def}_{\lambda\sigma\omega}(p,q,k)$ has mass dimension $3-d/2$
[see also the tree-level result in Eq.\ \pref{eq:gammatree}].
Thus, the quantity considered in Eq.\ \pref{eq:G3AAA} is inherently dimensionless
and we do not have to multiply it by any power of the lattice spacing $a$.
On the other hand, in order to get the corresponding
continuum quantities we have to multiply the momentum-space lattice gluon field
and the gluon propagator by $\sqrt{\beta}$ and by $\beta$, respectively. It follows that the
scalar function $G^{A^3}(p,q,\phi)$ has to be divided
by $\beta^{3/2}$ in order to obtain the corresponding continuum quantity.

Three different momentum configurations are shown in Fig.\ \ref{fg3v}.
The first is the momentum configuration used in
\cite{Parrinello:1994wd,Alles:1996ka,Cucchieri:2004sq,Boucaud:2000ey}
to study the behavior of the strong-coupling constant. Note that
this configuration is not well-defined in the continuum limit in case
the Gribov-Zwanziger scenario is correct, i.e.\ if the gluon propagator
vanishes at zero momentum. Of course, on a finite lattice this is not a problem.
In the second case, two momenta are equal and orthogonal.
Finally, the third case corresponds to three momenta with the
same magnitude. Furthermore, the behavior of $G^{A^3}(p,q,\phi)$ as a function
of $p$ and $q$ is shown in Fig.\ \ref{fg3vo}.

As said above, the statistical errors are quite large at large momenta, even after averaging
over more than $20,000$ configurations. Let us recall that, in some cases, the data
shown in Fig.\ \ref{fg3v} have been obtained by summing more than 700 different
terms. In particular we checked that, at large $p$, the central values of the data can change
substantially if one considers different subsets of the whole set of configurations used in
the analysis. This suggests that the statistical error obtained at large $p$
is likely under-estimated.

Despite the large statistical fluctuations at large momenta,
the vertex clearly decreases in the limit of small momenta.
This behavior can be observed in the three plots in Fig.\ \ref{fg3v}
and also in Fig.\ \ref{fg3vo}, for all directions approaching the
limit $p=q=k=0$.
On the other hand, at the smallest momentum point on the $30^3$ lattice at $\beta=4.2$
(see top panel in Fig.\ \ref{fg3v}), i.e.\ for the orthogonal configuration,
the vertex is clearly negative.
Of course, with our data we cannot say if the vertex would stay finite or
would become larger (in absolute value) as the zero momentum limit is approached.
Indeed, we know that, with our lattice volumes and $\beta$ values, the true infrared regime
is not reached yet, since the gluon propagator is not suppressed at small momenta
\cite{Cucchieri:2003di}.
Clearly, if the 3-point Green's function stays constant in the infrared limit, while
the propagators get suppressed, the vertex $\Gamma^{A^3\indexsep abc}_{\mu\nu\rho}(p,q,\phi)$ 
may be enhanced, as predicted by studies using functional methods in four dimensions \cite{Alkofer:2004it}
and in three dimensions \cite{Schleifenbaum:2006bq}.

Finally, in order to test whether other tensor structures are relevant,
the vertex has also been contracted with itself instead of using the tree-level vertex.
Results are not presented here, but we find that the corresponding scalar function
shows a strong increase compared to the tree-level behavior at large momenta. This suggests
that other tensor structures also contribute to the vertex in a non-trivial way.

\subsection{Ghost-gluon vertex}

The ghost-gluon vertex can be treated essentially along the same lines as
the three-gluon vertex. Let us recall that at the tree level, in the continuum, this
vertex is given by
\be
\Gamma_\mu^{\tl\indexsep c\bar cA\indexsep abc}(p,q,k)=igf^{abc}q_\mu ,
\label{eq:treelevcont}
\ee
while on the lattice one finds \cite{Rothe:1997kp}
\be
\Gamma_\mu^{\tl\indexsep L\indexsep c\bar cA\indexsep
              abc}(p,q,k)=igf^{abc} e^{i\pi k_{\mu}/N} \tilde{q}_\mu\cos(\hat{q}_\mu).
\label{eq:ccAlattice}
\ee
Again, the cosine and the exponential are lattice artifacts, going to 1 in the formal
continuum limit $a \to 0$.
Also, on the lattice it is only possible to determine the full Green's function
\be
G_\mu^{c\bar cA\indexsep abc}(p,q,k)\,=\,\frac{1}{V}<c^a(p)\bar c^b(q) A_\mu^c(k)> ,
\label{eq:Gmu}
\ee
where $c^a$ (respectively $\bar c^b$) is the ghost (respectively anti-ghost) field. Then, the
scalar quantity we evaluate is defined by
\begin{widetext}
\bea
G^{c\bar cA}(q,k,\phi)&=&\frac{\Gamma_\mu^{\tl\indexsep L\indexsep c\bar cA\indexsep abc}(p,q,k)\;
    G_\mu^{c\bar cA\indexsep abc}(p,q,k)}{\Gamma_\mu^{\tl\indexsep 
                            L\indexsep c\bar cA\indexsep abc}(p,q,k)\;
          D_G^{ad}(p)\;D_G^{be}(q)\;D_{\mu\nu}^{cf}(k)\;\Gamma_\nu^{\tl\indexsep
                            L\indexsep c\bar cA\indexsep def}(p,q,k)} 
\label{eq:Accfunction} \\[2mm]
&=&\frac{\Gamma_\mu^{\tl\indexsep L\indexsep c\bar cA\indexsep abc}(p,q,k)
     \;G_\mu^{c\bar cA\indexsep abc}(p,q,k)}{\Gamma_\mu^{\tl\indexsep
                            L\indexsep c\bar cA\indexsep abc}(p,q,k) \;
     P_{\mu\nu}(k) \; \Gamma_\nu^{\tl\indexsep L\indexsep c\bar cA\indexsep abc}(p,q,k)\;
                         D_G(p)\; D_G(q) \; D(k)} \\[2mm]
&=&\frac{-igf^{abc}\tilde{q}_\mu\cos(\hat{q}_\mu)\;
    G_\mu^{c\bar cA\indexsep abc}(p,q,k)}{g^2 N_c (N_c^2 - 1)
                  \tilde{q}_\mu\cos(\hat{q}_\mu)
      P_{\mu\nu}(k) \tilde{q}_\nu\cos(\hat{q}_\nu) \; D_G(p)\; D_G(q) \; D(k)} ,
\label{eq:ggAfunction}
\eea
\end{widetext}
where $D_G$ is the ghost propagator and we used  Eq.\ (\ref{eq:ccAlattice}).
Note that we have neglected in the denominator the color off-diagonal components
of the propagators and used the relation $f^{abc} f^{abc} = N_c (N_c^2 - 1)$.
Also, due to the implicit contraction of the full Green's function with a gluon propagator,
only one tensor structure survives, which is proportional to the incoming anti-ghost momentum $q$.
Finally, as for the three-gluon vertex, the exponential pre-factors cancel out in the numerator
and in the denominator of the scalar function defined above.

For the ghost-gluon vertex, the normalization factors are quite simple and will be given explicitly.
In the orthogonal case, we considered a gluon momentum $k$ aligned along the $y$-axis, while
the incoming anti-ghost momenta $q$ is chosen along the $x$-axis. This implies
$q_\mu P_\mn(k)=q_\nu$. Thus, the denominator in Eq.\ (\ref{eq:ggAfunction}) is proportional to
\be
\sum_\mu \tilde{q}^2_\mu \cos^2(\hat{q}_\mu) = \tilde{q}_x^2\cos^2(\hat{q}_x) ,
\ee
where the last equality follows because the momentum $q$ is aligned along the $x$-direction.
Clearly, in this case one recovers the kinematical normalization considered in Ref.\ \cite{Cucchieri:2004sq}.
In the case of equal momenta, the situation is more complicated and an explicit Gram determinant appears.
Indeed, by choosing the gluon momentum $k$ in the $x-y$ plane and the incoming anti-ghost momentum $q$
in the $x-z$ plane, the denominator is proportional to
\be
\tilde{q}_x^2\cos^2(\hat{q}_x)
   \left(1-\frac{\tilde{k}_x^2}{\tilde{k}_x^2+\tilde{k}_y^2}\right)
       +\cos^2(\hat{q}_z)\tilde{q}_z^2 \; .
\ee
In the case of equal components $\hat{q}_x = \hat{q}_z = \hat{k}_x = \hat{k}_y = \hat{q}$
this expression simplifies to $3\tilde{q}^2\cos^2(\hat{q})/2$.
Note that, in both cases, the normalization would be the same in the four-dimensional case, since
all momenta considered are in a three-dimensional sub-space.
Finally, it should be noted that the only Lorentz indices of $G_\mu^{c\bar cA\indexsep abc}(p,q,k)$
contributing to the numerator are those corresponding to non-zero components of the incoming
anti-ghost momentum $q$. Thus, in the orthogonal configuration the argument of the
exponential pre-factor appearing in Eq.\ (\ref{eq:Aofp}) is always zero for the  Lorentz components
contributing to $G_\mu^{c\bar cA\indexsep abc}(p,q,k)$.

\begin{figure*}
\includegraphics[width=\linewidth]{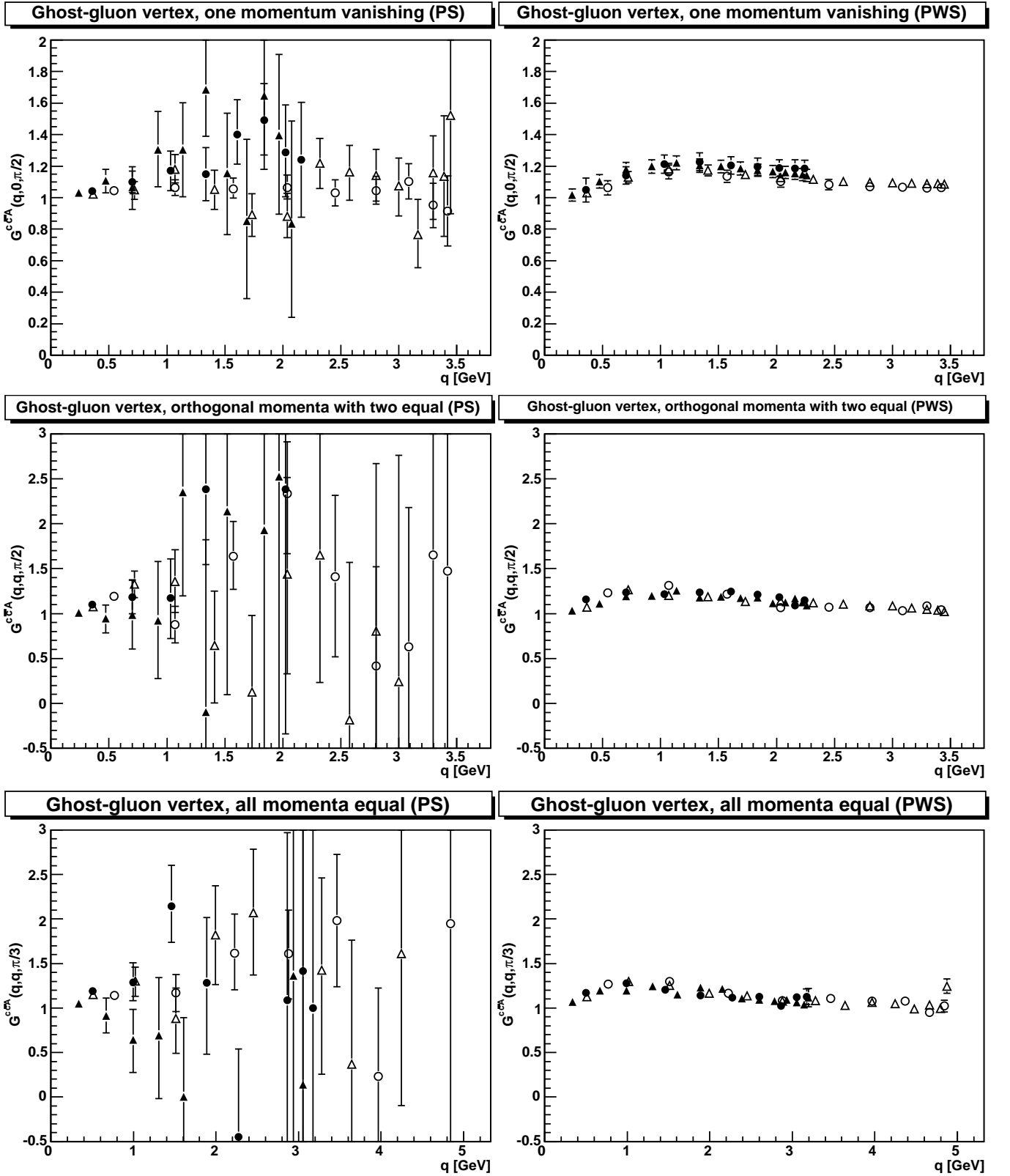}
\caption{\label{fggv} The scalar quantity $G^{c\bar cA}(q,k,\phi)$ defined in Eq.\
(\ref{eq:Accfunction}) as a function of the magnitude of the incoming anti-ghost momentum $q$.
Full symbols correspond to $\beta=4.2$ and open symbols to $\beta=6.0$;
circles are used for $V=20^3$ and triangles for $V=30^3$.
Results on the left-hand side have been obtained using a point source (PS) while on the
right-hand side we considered a plane-wave source (PWS).
In the top panel we show results for the orthogonal configuration with the gluon momentum $k$ vanishing.
In the middle panel we consider an orthogonal configuration with the two momenta ($q$ and $k$) having
the same magnitude.
In the bottom panel we plot data for the case with the three momenta equal.
For the number of configurations considered in the PWS case, see caption of Table \protect\ref{tab:eigen}.
}
\end{figure*}

\begin{figure}
\includegraphics[width=\linewidth]{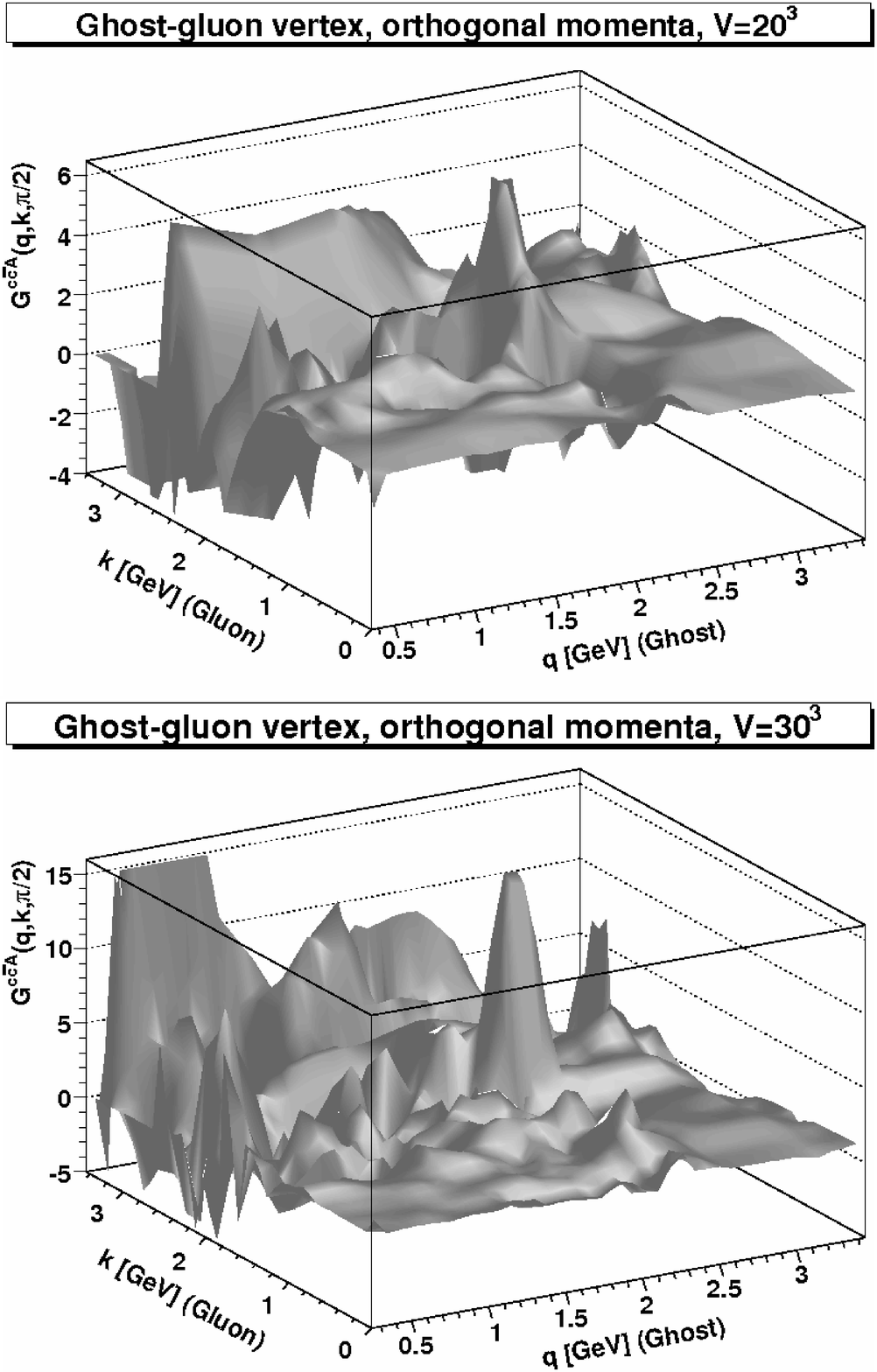}
\caption{\label{fggvo} The scalar quantity $G^{c\bar cA}(q,k,\phi)$ defined in Eq.\
(\ref{eq:Accfunction}) as a function of the magnitude of the
incoming anti-ghost momentum $q$ and of the gluon momentum $k$
(for the orthogonal configuration) using the point-source method.
Here, for each data point, we plot only the central value.
The data for $\beta=4.2$ and $\beta=6.0$ are plotted together
and they are interpolated by Gouraud shading, as implemented in the ROOT package \cite{Brun:1997pa}.
In the top figure we used data for the lattice volume $V=20^3$; in the bottom one
we consider the lattice volume $V=30^3$.
Spikes indicate positions where, due to fluctuations, the value is outside the drawing range.
The large spikes below the smallest non-zero anti-ghost momentum $q$ are an artifact of the interpolation.
}
\end{figure}

Let us also note that, when contracted with the transverse projector $P_{\mu\nu}(k)$ of the gluon momentum $k$
(as done here), the vertex should be invariant under the exchange of the ghost and of the
anti-ghost fields \cite{Lerche:2002ep}. This is of course true at the tree-level [see Eq.\ (\ref{eq:treelevcont})],
since it corresponds to an exchange of the color indices $a$ and $b$ and to replacing $q$ by $p=-k-q$.
Indeed, the term proportional to $k_{\mu}$ vanishes, due to the contraction with $P_{\mu\nu}(k)$,
and the minus sign of $-q_{\mu}$ cancels with the minus sign related to the antisymmetry of the
structure constant. In the general case this invariance is a consequence of a global $SL(2,R)$
symmetry between ghosts and anti-ghosts (see Appendix A in Ref.\ \cite{Alkofer:2000wg}
for a proof of this symmetry in Landau gauge).
As the Green's function \pref{gggv} is implicitly contracted with a gluon propagator, it follows that
\be
G_\mu^{c\bar cA\indexsep abc}(p,q,k)=G_\mu^{c\bar cA\indexsep bac}(q,p,k).
\ee
This result is also related to the symmetry of the Faddeev-Popov matrix under simultaneous
exchange of space and color indices.

The only remaining task is then to determine the quantity defined in Eq.\ (\ref{eq:Gmu}) above.
Since on the lattice one does not consider ghost (and anti-ghost) fields, the (equivalent)
expression to be used is \cite{Cucchieri:2004sq}
\be
G_\mu^{c\bar cA\indexsep abc}(p,q,k)= \frac{1}{V} < (M^{-1})^{ab}(p,q) \, A_\mu^c(k)>.\label{gggv}
\ee
One can employ the point-source method, used for evaluating the ghost propagator, also in the case of the
ghost-gluon-vertex function. Indeed, by using translational invariance we can write \footnote{Here,
in order to simplify the notation, we omit the color indices.}
\be
<M^{-1}(x,y)\, A(z)>\,=\,<M^{-1}(0,y-x)\,A(z-x)>  .
\ee
Then, by using momentum conservation, i.e.\ $p=-k-q$, and the equation above
we obtain
\bea
&&\frac{1}{V}\sum_{xyz}e^{2\pi i(-(k+q)x+qy+kz)/N} <M^{-1}(x,y) A(z)> \nn \\
&=&\frac{1}{V}\sum_{xyz} e^{2\pi i[q(y-x)+k(z-x))]/N}\nn\\
&& \qquad \qquad \times <M^{-1}(0,y-x) A(z-x)> \nn \\[3mm]
&=&\sum_{yz}e^{2\pi i(qy+kz)/N}<M^{-1}(0,y) A(z)> \nn \\
&=&\sum_{xyz}e^{2\pi i(qy+kz)/N}<M^{-1}(x,y) A(z) \,\delta_{x0}> \nn \\
&=&\sum_{xyz}e^{2\pi i(qy+kz)/N}<M^{-1}(x,y) A(z) (\delta_{x0} -\frac{1}{V})> . \nn
\eea
In the third line we have re-defined the indices $y$ and $z$ and summed over $x$.
In the last line we added a term that vanishes for non-vanishing ghost momenta
\cite{Boucaud:2005gg}.
Clearly, the quantity $\delta_{x0}-1/V$ has a null value when summed over $x$,
i.e.\ the inversion of the Faddeev-Popov operator is done in the subspace orthogonal to the
(trivial) kernel of $M$. Thus, one can evaluate the Green's function by considering the
Fourier-transformed gluon field and by inverting the Faddeev-Popov operator using the point-source
method. The final step requires one to evaluate the Fourier transform using the incoming anti-ghost momenta $q$, 
i.e.\ the one appearing in the tree-level vertex.
The third momentum (i.e.\ $p$) is implicitly defined by momentum conservation.
As said above, the point-source method has the advantage of only $N_c^2 - 1$ inversions per configuration
but, compared to the plane-wave-source method, requires many more configurations in order to
achieve a given statistical accuracy.

Note that, since the momentum-space ghost propagator $D_G$ has mass dimension
$-2$, we have that
the full Green's function $G_\mu^{c\bar cA\indexsep abc}(p,q,k)$
[see Eq.\ (\ref{gggv})] has mass dimension $-3-d/2$. At the same time,
the ghost-gluon vertex $\Gamma_\mu^{\tl\indexsep L\indexsep c\bar cA\indexsep abc}(p,q,k)$
[see Eq.\ (\ref{eq:treelevcont})] has mass dimension $3-d/2$.
Thus, also in this case the scalar function considered is clearly dimensionless
and we do not have to multiply it by any power of the lattice spacing $a$.
On the other hand, in order to get the corresponding
continuum quantities we have to divide it by $\beta^{1/2}$.

\begin{figure}
\includegraphics[width=\linewidth]{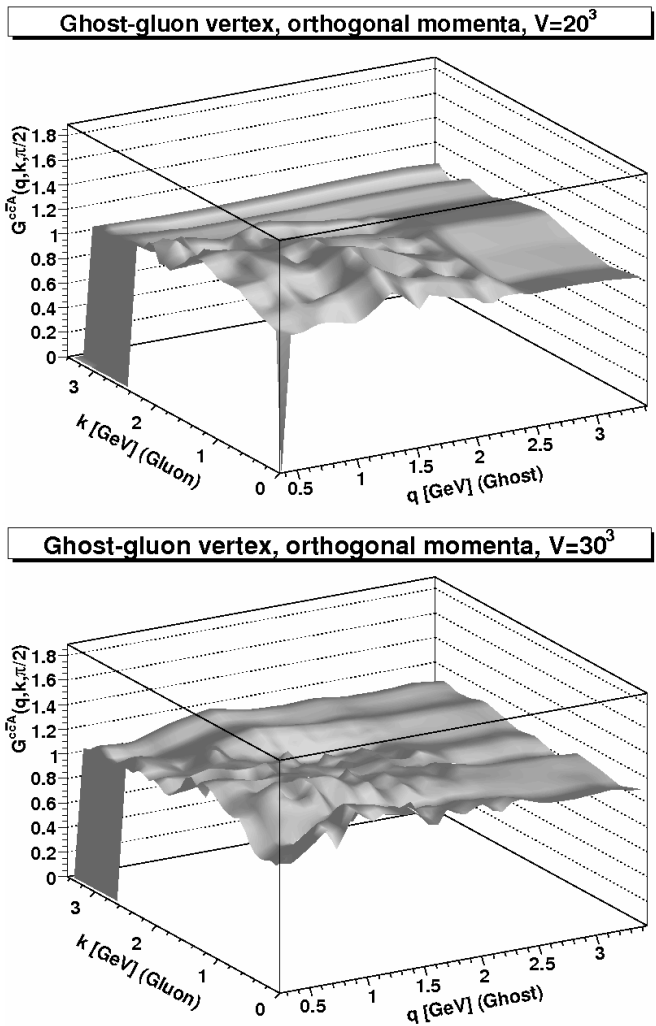}
\caption{\label{fggvo2} Same as Figure \ref{fggvo}, but using a mixed
point-source/plane-wave-source method. The number of configurations
considered here are 283 and 269 for the lattice volume $V = 20^3$ at
$\beta = 4.2$ and $\beta = 6.0$, respectively, and 227 and 213 configurations
for $V = 30^3$ at $\beta = 4.2$ and $\beta = 6.0$.
}
\end{figure}

Results are shown in Figs.\ \ref{fggv} and \ref{fggvo}.
In the left column of Fig.\ \ref{fggv} we show the results obtained using the point-source
method, while on the right column we present the data obtained using the plane-wave source,
for the vertex and for the ghost propagator appearing in Eq.\ \pref{eq:ggAfunction}.
We see agreement between the two sets of data but, as said above, the fluctuations are much
smaller in the latter case, the relative error being of the order of $5 \%$
for all momenta. On the other hand, when considering the point-source method (left column)
the fluctuations become rather large with increasing gluon momentum, i.e.\ the results
rapidly lose accuracy.
Let us finally note that the results on the right column do not change visibly if
one evaluates the ghost propagator using the point-source method and the vertex
using the plane-wave source (see Figure \ref{fggvo2}).

Clearly, in all cases
the vertex is essentially constant and of order one, for all momentum configurations,
confirming the results obtained in the 4d case \cite{Cucchieri:2004sq,Ilgenfritz:2006gp}.
This would imply a fulfillment of the corresponding Slavnov-Taylor identity \cite{Taylor:ff}.
This result is also in very good agreement with the functional predictions in the 3d case
\cite{Schleifenbaum:2006bq,Schleifenbaum:2004id} and the usual assumptions made in functional calculations.

\section{Summary and Outlook}\label{ssum}

In this work we have evaluated the
three-point vertices of (pure) Yang-Mills theory using
a wide range of momentum configurations. In order to reduce the computational
cost we considered the 3d $SU(2)$ case. We note that the data for the three-gluon vertex
suffer from large statistical errors at large momenta, even when considering
more than 20,000 configurations. Let us mention that, from a preliminary study \cite{prep},
this problem seems to be less severe in the 4d case. For the ghost-gluon vertex
the data also show a large ratio noise/signal if one uses the point-source
method. In this case the use of the plane-wave source (for the vertex)
allows to reduce the error below $10 \%$ already with a few tens of configurations.

As for the infrared behavior of the vertices, we found that
the three-gluon vertex becomes very small as the momentum decreases.
This mid-momentum suppression is similar to the behavior considered for this
vertex in Ref.\ \cite{Maas:2004se} 
in order to obtain a positive semi-definite gluon propagator.
From our data it is difficult to say what would be the behavior of this
vertex at very small momenta.
However, in one of the kinematical configurations considered here, the three-gluon vertex becomes
negative at the smallest non-zero momentum (about 240 MeV).
Thus, a possible scenario could be a vertex becoming larger (in absolute value)
as the zero momentum limit is approached. Let us note that a positive infrared divergent
three-gluon vertex has been recently found in functional studies
in three dimensions \cite{Schleifenbaum:2006bq}.

The ghost-gluon vertex stays constant and essentially
equal to the tree-level value in the range of momenta considered.
This is agreement with various theoretical predictions
\cite{Alkofer:2004it,Schleifenbaum:2006bq,Schleifenbaum:2004id} and with numerical results
for the 4d case \cite{Cucchieri:2004sq,Ilgenfritz:2006gp}.

Thus, our results, albeit exploratory, seem to support (at least at the
qualitative level) the Gribov-Zwanziger and Kugo-Ojima
scenarios of confinement and the central assumptions usually considered
in functional methods.
In particular, we have shown (see Figure \ref{fig:eigen}) that the smallest non-zero
eigenvalue of the Faddeev-Popov matrix goes to zero in the continuum limit, i.e.\
a (continuum) Landau configuration should belong to the first Gribov horizon.
On the other hand, these results should be taken with caution, as we know that in 3d
with the (physical) volumes considered here the true asymptotic infrared region has not
been reached yet \cite{Cucchieri:2003di}.

We plan to extend this study to larger lattice volumes, in order to explore the
far infrared limit. We also plan to consider other gauge groups, especially the physical $SU(3)$ group.
Finally, one should consider possible systematic effects related to the existence of
Gribov copies.

\acknowledgments

A.\ M. was supported by the DFG under grant number MA 3935/1-1.
A.\ C. and T.\ M. were supported by FAPESP (under grant \# 00/ 05047-5)
and by CNPq.

\end{document}